\documentclass[twocolumn,showpacs,preprintnumbers,amsmath,amssymb,footinbib]{revtex4-1}
\usepackage[utf8]{inputenc}
\usepackage[english]{babel}
\usepackage{lmodern}
\usepackage{amsmath}
\usepackage{graphicx}
\usepackage{amssymb}
\usepackage{epstopdf}
\usepackage{ upgreek }
\usepackage{color}
\usepackage{setspace}
\usepackage{bbm}
\usepackage[bottom]{footmisc}

\graphicspath{{Figures/}}

\usepackage[toc,page]{appendix}
\usepackage{tocloft}

\begin{document}

\title[]{Directional transport along an atomic chain}
\author{R.~Guti\'{e}rrez-J\'{a}uregui}
\email[Email:]{r.gutierrez.jauregui@gmail.com}
\author{A.~Asenjo-Garcia}
\email[Email:]{ana.asenjo@columbia.edu}
\affiliation{Department of Physics, Columbia University, New York, NY, USA.}

\date{\today}
\begin{abstract}
Motivated by a recent prediction to engineer the dispersion relation of a waveguide constructed from atomic components~[\textit{arXiv:2104.08121}], we explore the possibility to create directional transport in an open, collective quantum system. The optical response of the atomic waveguide is characterized through a scattering-matrix formalism built upon theories of photoelectric detection that allows us to find the required conditions for directional mode-to-mode transmission to occur and be measured in an experimental setting. We find that directional waveguides allow for an efficient outcoupling of light by reducing backscattering channels at the edges. This reduced backscattering is seen to play a major role on the dynamics when disorder is included numerically. A directional waveguide is shown to be more robust to localization, but at the cost of increased radiative losses.
\end{abstract}
\maketitle

An excited atom in free space will eventually find a way to its ground state.~In this spontaneous emission process, energy is originally localized inside a small volume from which it is set to travel outwards in the form of free photons~\cite{Fischer_2017,RGJ_2020}. When the atom is part of a dense and ordered atomic array, the excitation still finds a way out of the ensemble, but it does so through collective decay channels whose spatial and temporal profiles depend on the geometry of the array~\cite{Bettles_2016,Shahmoon_2017, Yoo_2020, Bloch_2020,Asenjo_2017_a}. The most simple example is that of a one-dimensional (1D) chain where an excitation can travel without losses until it finds an edge to escape through. These atomic arrays provide a versatile platform to study the controlled scattering of light in open, collective quantum systems whose response can be engineered and probed in real time~\cite{us_2021}. 

Such versatility can be used to generate directional transport along a 1D array by altering individual atomic constituents. Directional transport---where transmission is allowed in one direction and blocked in the other---has been at the center of intense research motivated in part to understand the motion of biological systems~\cite{Mateos_2000,Astumian_2002, Hanggi_2009}. Ratchet-type models predict that directional transport occurs when parity and time reversal symmetry are violated in otherwise unbiased source~\cite{Curie_1894}, and have been studied using elaborate atomic configurations where the internal degrees-of-freedom are used to generate periodic but asymmetric potentials to create directionality~\cite{Prost_1994,Rousselet_1994,Mennerat-Robilliard_1999}. These predictions have been supported by experimental observations using colloidal particles~\cite{Faucheux_1995}, polystyrene spheres~\cite{Rousselet_1994,Arzola_2017}, and cold rubidium atoms~\cite{Mennerat-Robilliard_1999}.

While these experiments describe the transport of material particles guided by an electromagnetic potential, an analogy is found in photonic systems where light is guided by matter~\cite{Hadad_2010,Jalas_2013,Sounas_2017}. The motivation behind directional transport in these platforms is to generate robust optical systems where backscattering is inhibited~\cite{MacKintosh_1988,Ortega_2018}. Through this constrain one can reduce the coupling to parasitic channels and, for imperfect materials, the interferences that give rise to localization~\cite{John_1984, Akkermans_1988,Lahini_2008}.

In this manuscript we present a systematic description of the transport of excitations along a directional atomic chain. We begin by reviewing an idealized model for an atomic chain whose optical response is engineered to display directionality, and calculate the transmittance of excitations via a scattering matrix. This approach is suitable to describe photons entering an atomic chain through a particular channel before leaving in another. We derive the conditions for directionality, and explore how to retrieve excitations efficiently from a directional chain. To finish, we include the effect of imperfections that break the periodicity of the array and show that backscattering is suppressed even in the presence of strong noise.  
%%%%%%%%%%%%%%%%%%%%%%%%%%%%%%%%%%
\section{Background: atomic chains}
%%%%%%%%%%%%%%%%%%%%%%%%%%%%%%%%%%

\begin{figure}[h]
\begin{center}
\includegraphics[width=.65\linewidth]{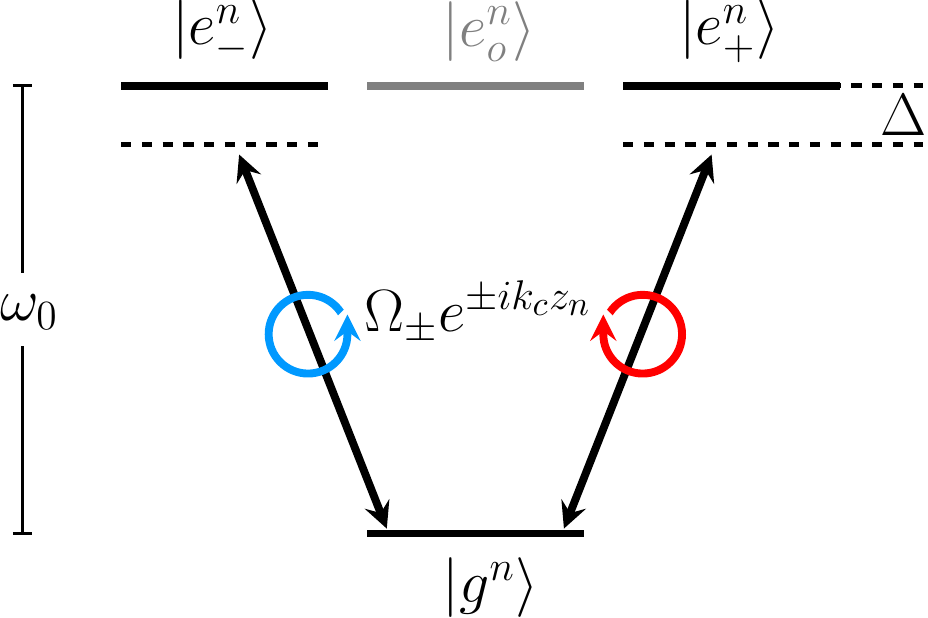}
\caption{Energy-level diagram to realize the effective Hamiltonian of Eq.~(\ref{eq:hamiltonian_effective}) where two excited states, denoted by $\vert e_{+}^{n} \rangle$ and $\vert e_{-}^{n}\rangle$, couple via a far-detuned Raman transition. Bold arrows represent the drive amplitudes $\Omega_{\pm}$ generated by two counter-propagating beams of wavevector $k_{c}$ and polarizations $\epsilon_{+}$ and $\epsilon_{-}$, such that the phase acquired from this two-photon process depends on the atomic position $z_{n}$.} \label{Fig:levels} 
\end{center}
\end{figure}

We consider an atomic chain made of $\mathcal{N}$ tightly trapped atoms separated a distance $a$. Each atom is characterized by its position $\mathbf{r}_{n}$ and is assumed to have a ground state $\vert g \rangle$ and three excited states $\vert e^{n}_{s} \rangle$ (connected to the ground states via photons of polarization $s =\{ 0,\pm\}$). States $\vert e^{n}_{+} \rangle$ and $\vert e^{n}_{-} \rangle$ are connected by a Raman transition as sketched in Fig.~\ref{Fig:levels}, where one leg of the transition is driven by a laser beam of amplitude $\Omega_{+}$ and phase $i k_{c} z_{n}$ (dependent on the atomic position) while the other is driven by a counter-propagating beam with amplitude $\Omega_{-}$ and phase $-i k_{c}z_{n}$. Both beams share the same frequency $\omega_{c} = k_{c}c$ and are far detuned from the atomic transition by $\Delta = \omega_{0} - \omega_{c}$. Their superposition defines a control field that distorts the atomic state. Under this configuration---and moving to an interaction picture with free Hamiltonian $\sum_{n,s} \hbar \omega_{c} \vert e^{n}_{s} \rangle \langle e^{n}_{s} \vert$---an effective Hamiltonian for the $n$th atom is realized~\cite{us_2021}:
\begin{align}\label{eq:hamiltonian_effective}
{\mathcal{H}}^{(n)} =& \sum_{s=\pm} \frac{\hbar}{2}(\Delta+\delta)({\sigma}^{(n)}_{ss}-{\sigma}^{(n)}_{gg} ) - \frac{\hbar\delta}{4}(1-s \cos \theta){\sigma}_{ss}^{(n)} \nonumber \\
+& \frac{\hbar \delta}{4} \sin \theta \left(e^{-2ik_{c}z_{n}}{\sigma}^{(n)}_{\scriptsize{+ -}} + e^{2ik_{c}z_{n}}{\sigma}^{(n)}_{\scriptsize{- +}} \right) \, .
\end{align}
Here ${\sigma}_{s s^{\prime}}^{(n)} = \vert e_{s}^{n}\rangle \langle e_{s^{\prime}}^{n}\vert$ is an operator connecting two atomic states; $\delta = (\Omega_{+}^{2} + \Omega_{-}^{2})/2\Delta$ is the light shift induced by the beams; and $\theta = 2 \arctan (\Omega_{+}/\Omega_{-})$ is a mixing angle. 

Atoms forming the chain interact with each other through the exchange of photons scattered in and out of the electromagnetic environment. In free space their dynamics can then be understood in terms of an open quantum system. By tracing the state of the electromagnetic field under the Born and Markov approximations, the master equation for the collective state of the atomic chain $\rho$ reads
\begin{equation}\label{eq:master_equation}
\dot{\rho} = \frac{1}{i\hbar} \left[\sum_{n,m} ( \mathcal{H}^{(n)} \delta_{nm} + \hbar\sum_{s,s^{\prime}} \Delta_{ss^{\prime}}^{nm} {\sigma}_{sg}^{(n)} {\sigma}_{gs^{\prime}}^{(m)} ), \rho \right] + \mathcal{L} [\rho],
\end{equation}
where $\mathcal{L}$ is the Lindblad superoperator 
\begin{equation}
\mathcal{L}[\cdot] = \sum_{n,m,s,s^{\prime}} \frac{\gamma_{ss^{\prime}}^{nm}}{2} \left( 2\sigma_{gs}^{(n)} \cdot \sigma_{s^{\prime}g}^{(m)} - \sigma_{s^{\prime}g}^{(m)} \sigma_{gs}^{(n)} \cdot - \cdot \sigma_{s^{\prime}g}^{(m)} \sigma_{gs}^{(n)}  \right),
\end{equation}
and the parameters $\Delta_{ss^{\prime}}^{nm}$, $\gamma_{ss^{\prime}}^{nm}$ represent the collective frequency shift and decay rate. These parameters depend on the relative position between two atoms $n$ and $m$ and their transition dipole moment via the electromagnetic Green's function of free space~\cite{green_us}. 

This chains supports the lossless transport of excitations via collective subradiant states generated by destructive interference of individual radiation paths. Subradiant states appear below a limiting lattice constant~\cite{Asenjo_2017}
\begin{equation}
a \leq \lambda_{0}/2 = \omega_{0}/4\pi c \, 
\end{equation}
and are characterized by vanishing eigenvalues of the collective decay matrix $\gamma_{ss}^{mn}$. As we show below, the subradiant channels can be engineered to be directional by changing the parameters of the effective Hamiltonian of Eq.~(\ref{eq:hamiltonian_effective}). This is a consequence of the control field that deforms the atomic dipole moment: first by creating an asymetric frequency shift that breaks the degeneracy between $\vert e_{s}^{\pm}\rangle$ states as an effective magnetic field would; and second by orienting the dipole moment in a spatially-dependent way and, in so doing, changing the way each atom of the array probes the local environment and its coupling to neighbouring sites. 

%%%%%%%%%%%%%%%%%%%%%%%%%%%%%%%%%%%%%%%%%%%%%%%%%%%%%%%%%%%%%
\section{Scattering matrix formalism}
%%%%%%%%%%%%%%%%%%%%%%%%%%%%%%%%%%%%%%%%%%%%%%%%%%%%%%%%%%%%%

Previous research on the transport through atomic arrays has been focused on the flux of excitations from one end of the chain to the other. Yet, when it comes to describe the light that enters and leaves the array, standard studies rely on physical intuition~\cite{Celardo_2009,Celardo_2010} or additional boundary conditions~\cite{Tsoi_2008,Asenjo_2017,Kornovan_2017,Mukhopadhyay_2019} that restrict the coupling to the edges of the array. This has proved to be a powerful tool to describe collective atomic systems but overlooks the spatial and temporal profile of the input and output fields that are ultimately measured in an experiment and can be problematic when discussing the mode-to-mode transmissions required for directional transport~\cite{Jalas_2013}. 

Here, we develop a scattering approach that describes the transport of excitations along an atomic chain. This method captures the absorption of a traveling photon by the chain and its ensuing emission into a desired channel. This approach is built from the theory of photoelectric detection~\cite{Carmichael_1999_b} and calculations for the scattering amplitudes of a photon by atomic systems~\cite{Cohen_1989}. In contrast to the master equation shown above, our focus now lies on free electromagnetic fields that, once detected, can be used to infer the emission path followed. 

To find the scattering matrix we take a step back and consider a system composed of the chain and its surrounding electromagnetic environment. These are described, respectively, by free Hamiltonians 
\begin{align}
\mathcal{H}_{S} &= \sum_{n} \mathcal{H}^{(n)} \, ,\\
\mathcal{H}_{R} &= \sum_{k, \lambda} \hbar \omega_{k} b^{\dagger}_{\mathbf{k},\lambda} b_{\mathbf{k},\lambda} \, ,
\end{align}
where $b_{\mathbf{k},\lambda}$ is the annihilation operator for a free electromagnetic mode of wavevector $\mathbf{k}$, frequency $\omega_{k}$, and polarization $\epsilon_{\mathbf{k},\lambda}$. These two subsystems couple through a dipolar term
\begin{equation} \label{eq:light_matter_coup}
\mathcal{H}_{SR} = \hbar \sum_{k, \lambda} \sum_{n,s}  \kappa_{\mathbf{k},\lambda}^{n,s} b^{\dagger}_{\mathbf{k},\lambda} \sigma_{gs}^{(n)} + \kappa_{\mathbf{k},\lambda}^{n,s *} b_{\mathbf{k},\lambda} \sigma_{sg}^{(n)},
\end{equation}
whose coupling parameter 
\begin{equation}	
\kappa_{\mathbf{k},\lambda}^{n,s} = \sqrt{\frac{\omega_{k}}{2\hbar\epsilon_{0} \mathcal{V} }} e^{-i \mathbf{k}\cdot \mathbf{r}_{n}} \epsilon_{\mathbf{k},\lambda} \cdot \mathbf{d}_{s}^{(n)} \, 
\end{equation}
illustrates how atoms probe the local amplitude of the electric field through their dipole moment~$\mathbf{d}_{s}$. For convenience we have considered a quantization volume $\mathcal{V}$ for the electromagnetic modes  that will later be taken to infinity. 

The free electric field operator is obtained by solving the Heisenberg equations of motion for the complete Hamiltonian $\mathcal{H} = \mathcal{H}_{S} + \mathcal{H}_{R} + \mathcal{H}_{SR}$ under the Born and Markov approximations. The resulting field separates into free and scattered fields~\cite{Cohen_1989,Carmichael_1999_a,Gardiner_1991}:
\begin{equation}\label{eq:total_field}
\mathbf{E}(\mathbf{R},t) = \mathbf{E}_{\text{free}}(\mathbf{R},t) + \mathbf{E}_{\text{scatt}}(\mathbf{R},t),
\end{equation}
where the positive frequency component of the latter is given in the far-field by
\begin{align}\label{eq:scattered_field}
\mathbf{E}^{(+)}_{\text{scatt}}(\mathbf{R},t) = \sqrt{\frac{\hbar \omega_{0}}{2 \epsilon_{o}c}  \frac{3 \Gamma_{0}}{8\pi}} &\sum_{n,s} \frac{\left[ 1 - (\mathbf{d}_{s}\cdot\mathbf{R}_{n})^{2} \right]^{1/2}}{R_{n}} \hat{\epsilon}_{ns}  \nonumber \\
&\times  e^{-i r_{n} k_{0}} \sigma_{gs}^{(n)} (t -R/c) \, .
\end{align}
Here $\mathbf{R}_{n} = \mathbf{R} - \mathbf{r}_{n} $ is the distance between the $n$th-atom and a point where the field is probed, while $\hat{\epsilon}_{ns}$ is an unitary vector pointing in the direction $\mathbf{R}_{n}\times ( \mathbf{R}_{n}\times\mathbf{d}_{s})$ that accounts for the radiation profile of each atom with an individual decay rate
\begin{equation}
\Gamma_0 = \frac{1}{4\pi \epsilon_0} \frac{4 \omega_{0}^{3} \vert d \vert^{2}}{3 \hbar c^{3}} \, .
\end{equation}

Through Eq.~(\ref{eq:scattered_field}) the scattered field can be used to probe the state of the atomic chain. Consider then a set of photodetectors surrounding the atomic chain that record all the photons being scattered into the environment. These detectors are placed at the positions $\mathbf{R}_{\theta\phi} = (R,\theta,\phi)$, with each one covering a surface area $R^{2} \Delta\Omega$ of solid angle $\Delta\Omega$ and considered capable of resolving the polarization state $\hat{\epsilon}_{\lambda}$ and arrival time of the photons. The information gained after each detection can be traced back to the state of the chain by applying the jump operator
\begin{align}\label{eq:jump_operator}
\mathcal{J}_{\theta\phi\lambda} &= \sqrt{\frac{2 \epsilon_0 c }{\hbar \omega_{0}}(R^{2} \Delta \Omega)} \mathbf{E}^{(+)}_{\text{scatt}}(\mathbf{R}_{\theta\phi})\cdot \hat{\epsilon}_{\lambda} 
\end{align}
and accounting for the necessary free evolution. These operators have units of square root of photon flux, such that
\begin{equation}\label{eq:probability}
P_{\theta\phi\lambda} = \tau\text{Tr}_{\mathcal{S}}\left[\mathcal{J}_{\theta\phi\lambda} \rho \mathcal{J}_{\theta\phi\lambda}^{\dagger} \right]
\end{equation}
gives the probability for a chain in state $\rho$ to scatter into the detector $(\mathbf{R}_{\theta, \phi}, \hat{\epsilon}_{\lambda})$ during a small time interval $\tau$. The trace is taken over atomic variables only.

Equations~(\ref{eq:scattered_field})-(\ref{eq:probability}) give the basic tools to unravel the state of the atomic chain subject to a particular measurement record and recover the path an excitation followed across the chain~\cite{Carmichael_1999_b}. We, however, are not interested in the particular times at which an input photon enters and an output photon leaves an otherwise empty chain; but in the probability amplitude for the process to take place. A sum over all the records where this process took place is given by the scattering $\mathcal{S}$-matrix whose components $\mathcal{S}_{ba} = \langle g;b \vert \mathcal{S} \vert g;a \rangle$ give the probability amplitude for a free field of energy $E_{a}$ and state $\vert a\rangle$ (\textit{e.g.}, $\vert \mathbf{k}_{a},\lambda_{a}\rangle$) to scatter into one of energy $E_{b}$ and state $\vert b \rangle$. In the reciprocal space these components are written as~\cite{Cohen_1989}: 
\begin{equation}\label{eq:scattering_matrix_initial}
\mathcal{S}_{ba}(E_{a}) = \delta_{\lambda_{a},\lambda_{b}} \delta(\mathbf{k}_{a}-\mathbf{k}_{b}) - 2\pi i \delta(E_{a}-E_{b}) \mathbf{T}_{ba}
\end{equation}
with
\begin{align}\label{eq:transition_matrix_initial}
&\mathbf{T}_{ba}= \\\nonumber
&\sum_{ns,ms^{\prime}} \langle g;0 \vert \left(\hbar \kappa_{\mathbf{k}_{b},\lambda_{b}}^{n,s *} \sigma_{gs}^{(n)} \right) \mathcal{Q} \frac{1}{E_{a} - \tilde{\mathcal{H}}} \mathcal{Q} \left(\hbar \kappa_{\mathbf{k}_{a},\lambda_{a}}^{m,s^{\prime}} \sigma_{s^{\prime}g}^{(m)} \right)  \vert g;0 \rangle \, .
\end{align}
This transmission matrix $\mathbf{T}$ divides environment and chain by connecting free fields to spin-waves through the operator $\mathcal{Q}=\sum_{n,s}\vert e^{n}_{s};0 \rangle \langle e^{n}_{s};0 \vert$, a projector into the subspace where one excitation populates the chain and the state of the field is vacuum. Once in this subspace, the resolvent $G(E) =  (E - \tilde{\mathcal{H}})^{-1}$ determines the channels the excitation can follow. This is done through a non-Hermitian Hamiltonian 
\begin{equation}\label{eq:non_hermitian}
\tilde{\mathcal{H}} = \mathcal{H}_{S} -  \sum_{n,m=1}^N \sum_{s=\pm} \hbar(\Delta_{ss}^{nm} + i \gamma_{ss}^{nm}) \hat{\sigma}_{sg}^{(n)} \hat{\sigma}_{gs}^{(m)} \, ,
\end{equation} 
that acts over atomic states only and displays the collective frequency shifts and decay rates caused by their self-consistent interaction with the environment [in connection to the master equation, Eq.~(\ref{eq:master_equation})].

We now bring together the picture provided by the jump operators of Eq.~(\ref{eq:jump_operator}) and the $\mathcal{S}$-matrix of Eq.~(\ref{eq:scattering_matrix_initial}) to study the transport along the chain. The key point is that the atomic ensemble only responds to free-field modes whose frequencies are close to the atomic resonance frequency $\omega_{0}$. For these frequencies, we can write~\cite{Carmichael_1999_a}
\begin{equation}\label{eq:jumps_decay}
\sum_{ns}\kappa_{k_{b},\lambda_{b}}^{ns*} \sigma_{gs}^{(n)} = \frac{1}{\sqrt{2\pi g(\omega_{0})}} \mathcal{J}_{p_{b}q_{b}\lambda_{b}}
\end{equation}
where $g(\omega_{0}) = R/6\pi c$ is the optical mode density at the atomic transition frequency. Thus, after integrating Eq.~(\ref{eq:scattering_matrix_initial}) over a small range of output modes $(N,N+dN)$ with $d{N} = g(\omega_{0})d\omega_{b}$, the $\mathcal{S}$-matrix takes the form
\begin{equation}\label{eq:scattering_matrix_ri}
\mathcal{S}(E) = \mathbbm{1} - i \textbf{t}(E)
\end{equation}  
where $\mathbbm{1}$ is the identity matrix and the transmission matrix
\begin{equation}\label{eq:transmision_matrix_ri}
\textbf{t}(E) = \left(\sum_{\beta} \mathcal{J}_{\beta} \right) \frac{1}{E - \tilde{\mathcal{H}}} \left(\sum_{\alpha} \mathcal{J}^{\dagger}_{\alpha} \right) \, 
\end{equation}
accounts for all input and output modes through a sum of jump operators $\mathcal{J}_{\beta}, \mathcal{J}^{\dagger}_{\alpha}$ that runs over all detectors $(\theta, \phi)$. This transmission determines the channels through which a photon enters the chain, propagates across it, and then scatters out. 

Equations~(\ref{eq:scattering_matrix_ri}) and~(\ref{eq:transmision_matrix_ri}) describe the main result of this section. They present a contextual description for the transport of excitations where input and output channels are given by the jump operators~$\mathcal{J}^{\dagger}_{\theta\phi\lambda}$ and~$\mathcal{J}_{\theta\phi\lambda}$. And, while developed with a scattering picture in mind, these equations can be written in a form that is more suitable for transport by choosing a different set of jump operators. We could choose, for example, the jump operators given by the eigenvectors of the collective decay matrix $\gamma_{ss}^{nm}$ denoted here by $\vert \phi^{(\nu)} \rangle = \sum c_{ns}^{(\nu)} \vert e_{s}^{n} \rangle$ with eigenvalues $\gamma_{\nu}$. Under this unraveling the jump operators take the form
\begin{equation}\label{eq:jumps_normal_modes}
\mathcal{J}_{\nu} = \sqrt{\gamma_{\nu}} \sum_{n,s} c_{ns}^{(\nu)} \sigma_{gs}^{(n)} \, .
\end{equation}

Normal mode and physical space representations are connected through the equality
\begin{equation}
 \sum_{n,m,s} \gamma^{nm}_{ss} \sigma_{gs}^{(n)}  \cdot \sigma_{sg}^{(m)} = \sum_{\alpha} \mathcal{J}_{\alpha} \cdot \mathcal{J}_{\alpha}^{\dagger} \, ,
\end{equation}
where $\alpha$ runs along $\nu$ or $\{\theta, \phi, \lambda\}$ to select a representation. For $\alpha = \{\theta, \phi, \lambda\}$ the right-hand side describes fields measured at particular points while for $\alpha = \nu$ it focuses on fields radiated by the normal modes, which have only a formal meaning. Notice that both sets of jump operators guarantee a unitary $\mathcal{S}$-matrix since 
\begin{equation}
\tilde{\mathcal{H}} - \tilde{\mathcal{H}}^{\dagger} = \sum_{\alpha} \mathcal{J}_{\alpha}^{\dagger} \mathcal{J}_{\alpha} \, .
\end{equation} 

A similar formula for the $\mathcal{S}$-matrix has been used to study nuclear reactions~\cite{Mello_1979,Sokolov_1992} and has also emerged in the context of mesoscopic systems~\cite{Celardo_2009, Celardo_2010} where focus is placed on the resonance spectra and its relation to its transport properties~\cite{Dittes_2000} rather than on the connection to scattering records. 

%%%%%%%%%%%%%%%%%%%%%%%%%%%%%%%%%%%%%%%%%%%%%%%%%%%%%%%%%%%%%%%%%%%%%%%%
\section{Transport in directional chains}
%%%%%%%%%%%%%%%%%%%%%%%%%%%%%%%%%%%%%%%%%%%%%%%%%%%%%%%%%%%%%%%%%%%%%%%%

The scattering matrix is now used to analyze the transmission across an atomic chain with emphasis on its directionality. This is done by considering the system Hamiltonian of Eq.~(\ref{eq:hamiltonian_effective}) and the normal mode representation of Eq.~(\ref{eq:jumps_normal_modes}) (jump operators with $\alpha = \nu $). We begin by shifting our attention from the scattered field towards the atomic states using Eqs.~(\ref{eq:scattering_matrix_ri}) and~(\ref{eq:transmision_matrix_ri}) whose components $\mathcal{S}_{ns;ms^{\prime}}$ represent the probability amplitude for a photon to enter the chain through the single atom state $\vert e_{s^{\prime}}^{m}\rangle$ and leave through $\vert e_{s}^{n}\rangle$ disregarding the photonic spatial profile.

An optical medium presents directionality when the propagation of excitations along two opposing paths displays different mode-to-mode transmissions~\cite{Jalas_2013}. This occurs when reciprocity is broken, a condition that is represented by an asymmetric scattering matrix such that
\begin{equation}
S_{ns;ms^{\prime}}(E) \neq S_{ms^{\prime};ns}(E)  \, .
\end{equation}
For the atomic chain described above, reciprocity is broken when collective decay and free operators do not commute
\begin{align}\label{condition_NR}
\left[\mathcal{H}_{S}, \sum_{n,m=1}^{\mathcal{N}} \sum_{s=-1}^{1} \gamma_{ss}^{nm} \sigma_{sg}^{(n)} \sigma_{gs}^{(m)} \right] \neq 0 \, .
\end{align}
This condition is satisfied for $\theta \neq n \pi$ and $\omega_{c} z_{n}/c \neq n \pi/2 $ for all $n$. The first requirement leads to an asymmetric frequency shift of $\vert e_{+} \rangle$ and $\vert e_{-} \rangle$ states, an effective Zeeman shift created from the atomic response to the elliptic polarization of the control field. The second requirement corresponds to a subwavelength rotation of the atomic dipoles that is generated from the polarization gradient of the same field. These two requirements---simultaneous time-reversal and parity symmetry breaking---were found to be necessary for a waveguide made from plasmonic particles to break reciprocity and display directionality~\cite{Hadad_2010}. Equation~(\ref{condition_NR}) formalizes this result and extends it for an atomic chain.

\begin{figure}[h]
\begin{center}
\includegraphics[width=1.\linewidth]{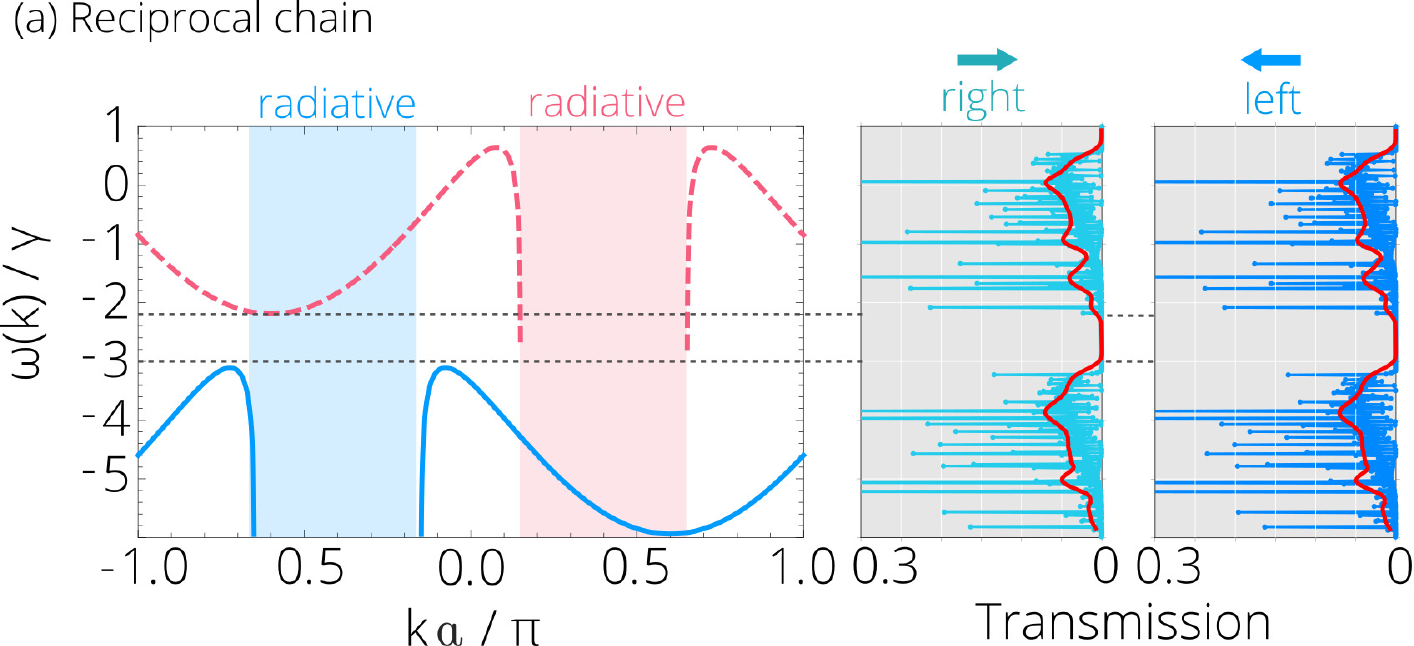}
 \includegraphics[width=1.\linewidth]{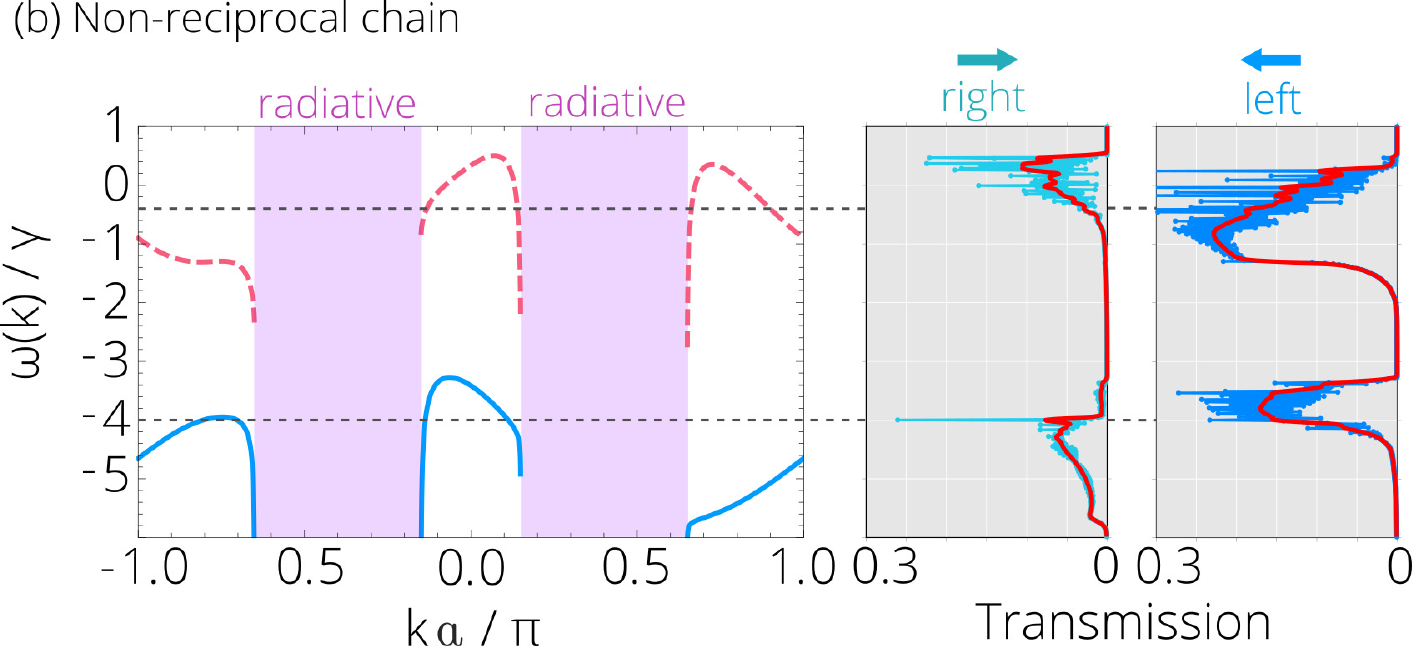}
\caption{Comparison between the dispersion relation for an infinite chain (left) and transmission of a finite chain of $\mathcal{N}=205$ atoms (right) under conditions of reciprocity and non-reciprocity. The dispersion relation marks a transparency window composed of two branches given by the frequencies of $\vert \text{l},k \rangle$ (dashed pink) and $\vert \text{u}, k\rangle$ (solid blue) states of Eq.~(\ref{eq:eigenstates}). The transmittance is obtained from $\sum_{s,s^{\prime}}\vert \langle n,s \vert\mathcal{S}\vert m,s^{\prime} \rangle \vert^{2}$ [see Eq.~(\ref{eq:scattering_matrix_ri})] with $n,m$ denoting the emission from atoms at the end of the chain. A finite transmittance is found for incoming photons whose energy matches the narrow resonances of the subradiant modes that fill the transparency window; as shown by gray dashed lines. Red lines in the transmission indicate the average over a small energy interval to visualize the infinite chain limit. For both plots the lattice constant is $a=\lambda_0/8$ and the Raman channels of Eq.~(\ref{eq:hamiltonian_effective}) have a strength $\delta= 10\Gamma_{0}/3 $ and phase $k_{c} = \pi/5a$ with $\theta=0$ in (a) and $\theta=\pi/4$ in (b).} \label{Fig:transmission} 
\end{center}
\end{figure}

Figure~\ref{Fig:transmission} shows the transmittance as a function of the input photon frequency for a chain of $\mathcal{N}=205$ atoms under conditions of reciprocity~(Fig.~\ref{Fig:transmission}a) and non-reciprocity~(Fig.~\ref{Fig:transmission}b). The transmittance is given by $\sum_{s,s^{\prime}}\vert \langle n,s \vert\mathcal{S}\vert m,s^{\prime} \rangle \vert^{2}$ with $n=1$ $(\mathcal{N})$ and $m = \mathcal{N}$ $(1)$, which gives the probability for a photon to be absorbed by an atom at one end of the chain and be emitted at the opposite end. The transmittance of a right- (light green) and left-propagating (blue) excitation displays an imbalance when reciprocity is broken. In both cases transmission channels appear as narrow resonances due to the atom-atom interactions~\cite{Fano_1961}. As more atoms are added to the chain additional resonances with a narrowing width begin to appear, generating a broad transparency window where excitation can propagate without losses. We have considered atoms at the edges since subradiant modes tend to scatter out of the chain at these points. While not shown in the figure, the transmittance is reduced for atoms separated from the edge as they are more likely to absorb photons through short-lived superradiant channels. 

Figure~\ref{Fig:transmission} also shows the dispersion relation of the atomic chain formed from the subradiant states. The dispersion relation is obtained from the infinite chain limit by diagonalizing the non-Hermitian Hamiltonian~(\ref{eq:non_hermitian}) as done in Ref.~\cite{us_2021}. In this scenario subradiant states---and their frequencies---are characterized by a wavevector $k$ directed along the chain axis and determined by the lattice separation, and a polarization index $\{ \text{u},\text{l}\}$. Written within the free basis these states read
\begin{subequations}\label{eq:eigenstates}
\begin{align}
\vert {\text{{u}}}, k \rangle & = \sum_{n} e^{i k z_{n}} \sum_{s=\pm}\left[ e^{-i s k_{c}z_{n}} c_{\text{u},s}^{(n)} \,\hat{\sigma}_{sg}^{(n)} \right]  \vert g \rangle^{\otimes \mathcal{N}} \, , \\
\vert {\text{{l}}}, k \rangle & = \sum_{n} e^{i k z_{n}} \sum_{s=\pm}\left[ e^{-i s k_{c}z_{n}} c_{\text{l},s}^{(n)} \,\hat{\sigma}_{sg}^{(n)} \right]  \vert g \rangle^{\otimes \mathcal{N}}  \, .
\end{align}
\end{subequations}
Here, the spatial phase $k_{c}z_{n}$ is inherited from the control field of Fig.~\ref{Fig:levels} and the probability amplitudes $c_{\text{u}/\text{l},s}^{(n)}$ are given explicitly in Ref.~\cite{us_2021}. The dispersion relations of these two branches are shown in Fig.~\ref{Fig:transmission} as dashed pink lines for the lower $\text{l}$-branch and solid blue line for the upper $\text{u}$-branch. In this sense, the dispersion relation provides a geometrical picture for the density of states and, through its derivative, the group velocities of traveling spin-waves. These subradiant states can be understood as guided modes that propagate along the chain wihtout scattering. They are obtained for quasimomentum lying beyond the light line $\vert k \pm k_{c} \vert > \omega_{0}/c$, thus allowing for radiation paths between different atoms to interfere destructively and cancel out the collective radiation, and generate a transparency window for lossless propagation whose width scales with the atomic separation as~$\sim\Gamma_{0}/(k_{0} a)^{3}$. As shown in Fig~\ref{Fig:transmission} the transparency window obtained by the scattering matrix method coincides with that generated by the dispersion relation of the subradiant states.  When reciprocity is broken, the dispersion relation becomes asymmetrical and counter-propagating channels display different transmittances. 

%%%%%%%%%%%%%%%%%%%%%%%%%%%%%%%%%%%%%%%%%%%%%%%%%%
\subsection{Emission from a traveling spin-wave}
\begin{figure}[h]
\begin{center}
\includegraphics[width=1.\linewidth]{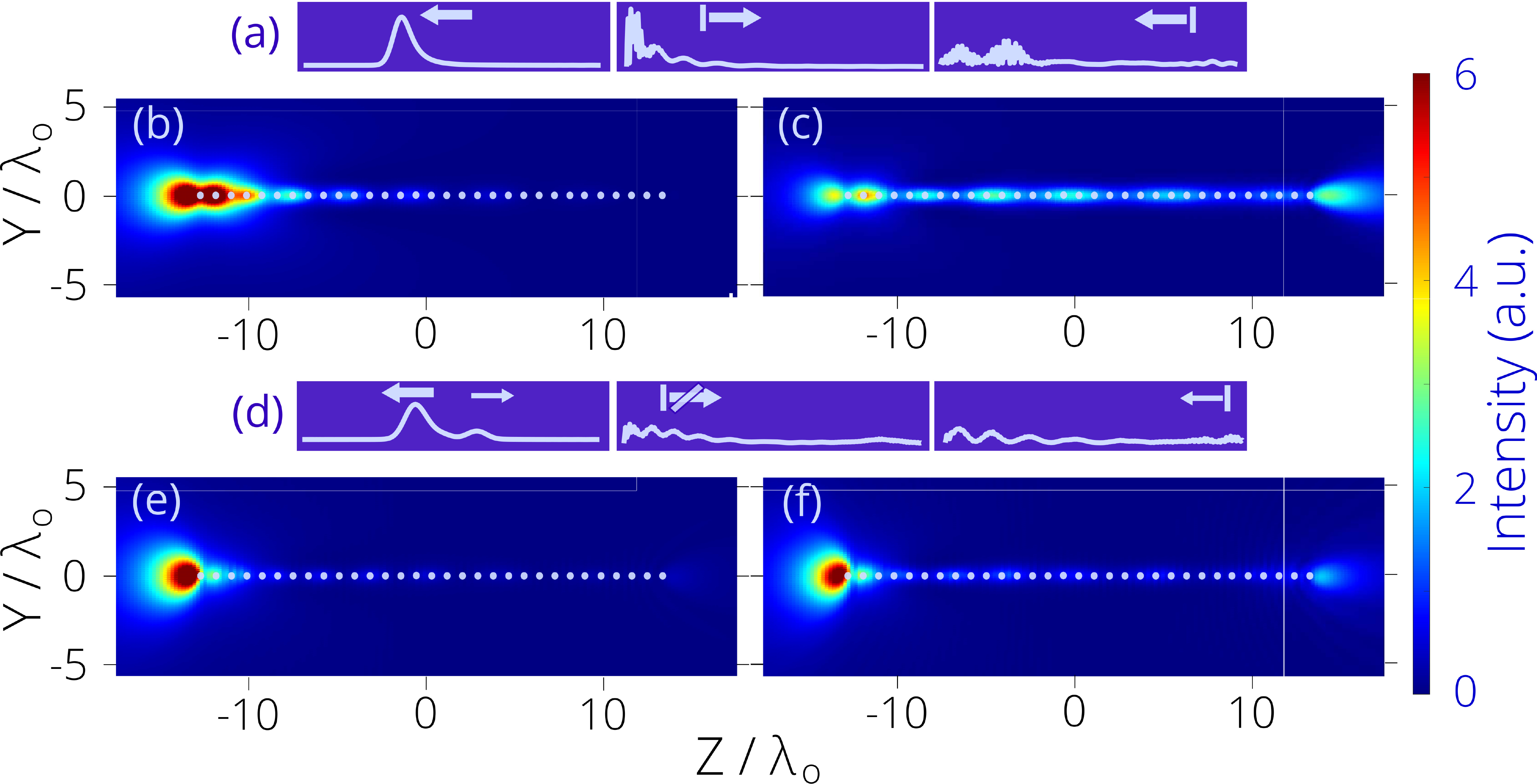}
\caption{Backscattering from a spin-wave bouncing off the ends of a chain under conditions for regular (a-c) and directional transport (d-f) used in Fig.~\ref{Fig:transmission}. Panels (a) and (d) represent the atomic population before~(left) and after the first~(center) and second bounces~(right). Notice the interference that arises in the central panel of the regular chain and is missing in the directional chain. Through Eq.~(\ref{eq:spin_wave}) we have decided to excite two different-frequency modes of the directional chain to exemplify how the backscattering is completely supressed for the left-propagating mode while reduced for the right-propagating one. The scattered field intensity at first [(b) and (e)] and second [(c) and (f)] bounces shows the inbalance in left and right transmissions of a directional chain that can be used to route light efficiently.} \label{Fig:scattered_field} 
\end{center}
\end{figure}

%%%%%%%%%%%%%%%%%%%%%%%%%%%%%%%%%%%%%%%%%%%%%%%%%%
The radiation paths are not cancelled completely for finite arrays, thus coupling subradiant modes to the environment and to each other. The backscattering into other modes, however, can be inhibited for directional chains and light can be routed into a given direction. This is exemplified in Fig.~\ref{Fig:scattered_field} where we plot the field intensities of a field scattered by a spin-wave inside the chain under conditions of regular and directional transport. We illustrate the different behaviors by considering a single-excitation initial state
\begin{equation}\label{eq:spin_wave}
\vert \psi \rangle = c_{g} \vert g \rangle^{\otimes \mathcal{N}} + c_{e} \sum_{n} \frac{a}{\sqrt{2\pi} \Delta x } e^{ia(k + k_{c}) - \frac{a^{2}(n-n_{o})}{\Delta x^{2}}} \vert e^{n}_{-} \rangle  \, ,
\end{equation}
that is set to evolve under the non-Hermitian Hamiltonian of Eq.~(\ref{eq:non_hermitian}). The chain is weakly populated $(\vert c_{e}\vert^{2} =0.2)$ and the excitation is centered around the site $n_0=100$ with spatial width $\Delta x^{2} = 60a^{2} $ and quasimomentum $k=0$. The evolution of this spin-wave is sketched in panels (a) and (d) where the atomic population is plotted at three different times: (i) before the wave reaches the end of the chain, (ii) as it reaches this end and is backscattered, and (iii) as the backscattered wave reaches the opposite end. The scattered field intensity at times (ii) and (iii) is plotted, respectively, at panels (b) and (c) for the regular chain and in (e) and (f) for the directional chain, following Eq.~(\ref{eq:scattered_field}). In both cases the intensity of the field is concentrated at one end of the chain as the spin-wave is bounced of the edge [(b) and (e)], but, with imbalanced backscattering channels, it spreads significantly different afterwards [(c) and (f)]. A spin-wave bouncing off one end of a regular chain backscatters into several subradiant channels that guide it to the opposite end. This is suggested by the scattered field intensity and the interference profile in the atomic population. In a directional chain, by contrast, the spin-wave has fewer channels to backscatter into and the emission remains localized off one end, as suggested by the population profile. The scattered field remains trapped on one side and the population interference pattern is lost.

Figure~\ref{Fig:scattered_field} shows two different spin-waves propagating along the directional chain. These counter-propagating waves are created from the initial state that overlaps with both excitation branches, $\vert \text{u} \rangle $ and $\vert \text{l} \rangle$, defined in Eq.~(\ref{eq:eigenstates}) since both polarizations are coupled through the Raman channels responsible of the directional response. This helps to illustrate the difference when backscattering channels are inhibitted or completely absent. In the case of the u-mode the spin-wave can find states to backscatter, thus leading to a low-intensity field travelling along the chain and a reduced interference pattern on the atomic population. Both these properties are reduced for the l-mode.

%%%%%%%%%%%%%%%%%%%%%%%%%%%%%%%%%%%%%%%%%%%%%%%%%
\subsection{Effect of disorder}
%%%%%%%%%%%%%%%%%%%%%%%%%%%%%%%%%%%%%%%%%%%%%%%%%
Throughout the last two sections we have joined together the dynamics inside an atomic chain to the scattered field with the objective of studying the transport properties of a directional chain and their relation to the far-fields measured in an experimental setting. We have emphasized the role of subradiant states that guide an excitation from one end of the chain towards the other through loss-less collective channels. With subradiant states emerging from the phase coherence between individual atomic constituents, the question remains as to how the transport of excitations is affected by imperfections of the array. 

Imperfections can manifest in our model through individual frequency shifts caused by the trapping potential or displacements in the atomic positions due to weaker traps. The effect in both cases is to break the periodicity of the array. We introduce these imperfections below and compare the response between reciprocal and non-reciprocal chains, showing that the transport properties of the latter are more resilient to disorder. 

We focus on individual frequency shifts for simplicity. They are given by an additional potential 
\begin{equation}
V = \sum_{n,s} \mathcal{E}_{n} \sigma_{ss}^{(n)} \, ,
\end{equation}
where $\mathcal{E}_{n}$ is a stochastic variable distributed over a frequency band of zero mean and variance $\sqrt{W}$,
\begin{equation}
\langle \mathcal{E}_{n} \rangle_{\text{avg}} = 0 \, ; \,  \langle \mathcal{E}_{n} \mathcal{E}_{m} \rangle_{\text{avg}} = W \delta_{nm} \, .
\end{equation}
Since we are interested in the effect over atomic coherence it is convenient to write this potential in the reciprocal space where
\begin{equation}
V = \sum_{n,s} \sum_{k,k^{\prime}} \mathcal{E}_{n} e^{i(k-k^{\prime})z_{n}} \vert e_{s},k \rangle \langle e_{s}, k^{\prime} \vert \, ,
\end{equation}
as obtained from the relation $\langle k \vert e_{s}^{(n)} \rangle = e^{i k z_{n}} \vert e_{s} \rangle$. A similar decomposition can be done for random atomic positions with the added complexity that the interaction strength can diverge for small lattice sites.

The role of this imperfection is to couple states of different quasimomentum $k$, causing a state of well-defined wavevector, \textit{e.g.}, a spinwave or a normal mode, to spread in reciprocal space and localize in position. For weak energy shifts whose variance is significantly smaller than the transparency window, the impurities can be treated as a stochastic disorder that deform the dispersion relation by coupling states of approximately the same energy. Ultimately, this coupling reduces the atomic coherence with a more pronounced effect over frequencies with a high density-of-states~\cite{Lahini_2008}. 
\begin{figure}[h]
\begin{center}
\includegraphics[width=1.\linewidth]{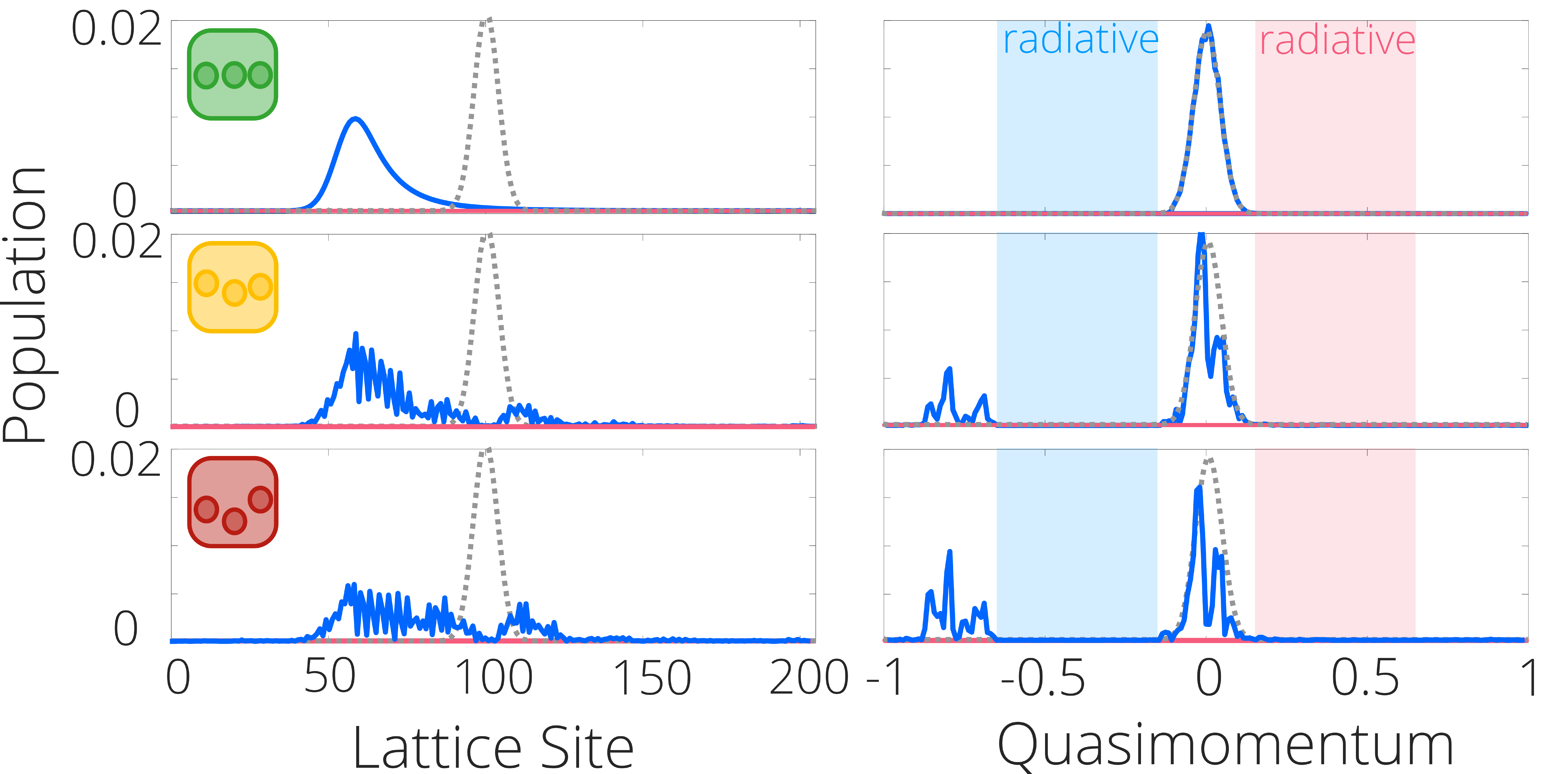}
\includegraphics[width=1.\linewidth]{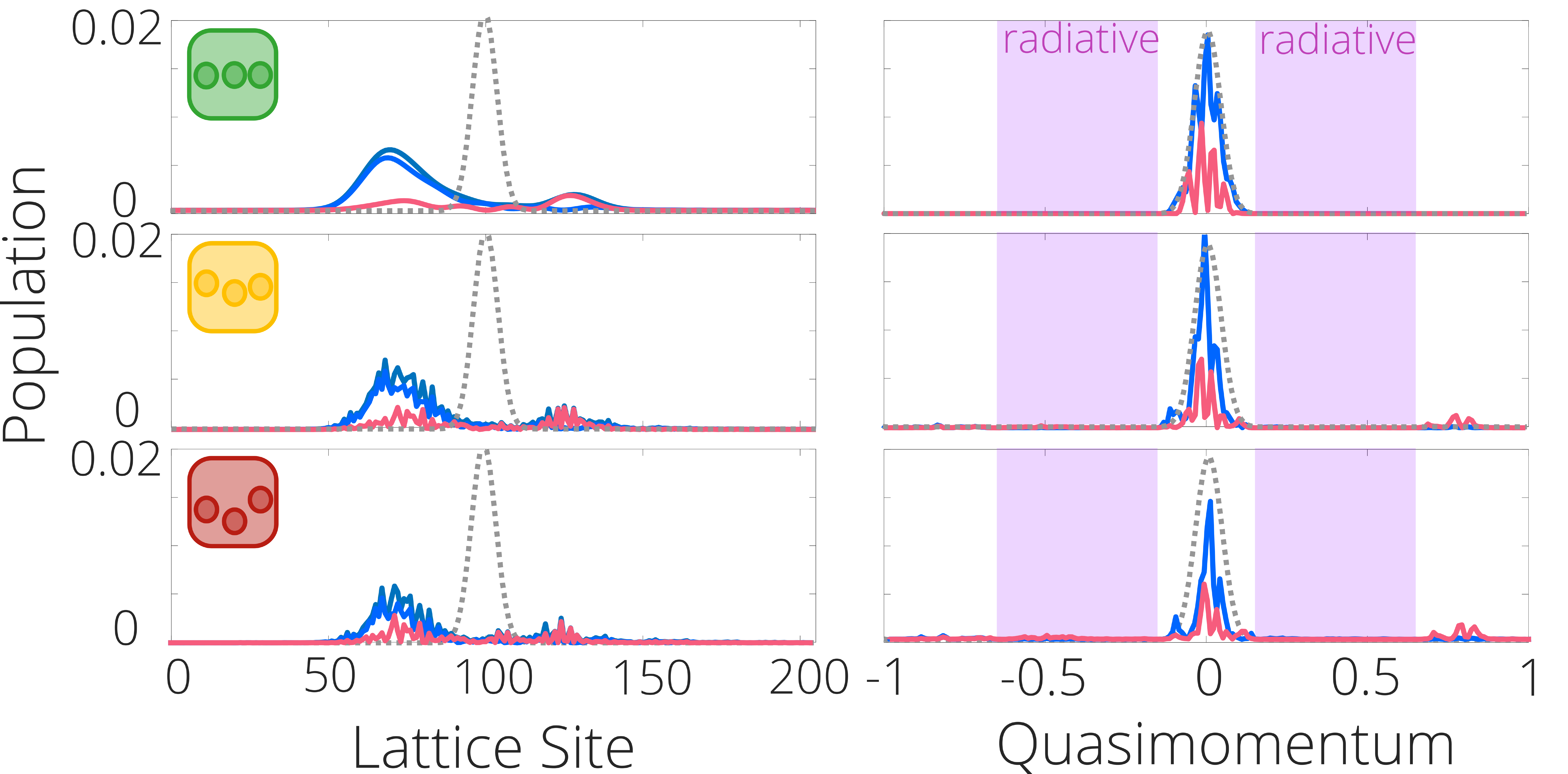}
\caption{Transport of a spinwave in the presence of disorder for the regular and directional chains of Fig.~\ref{Fig:transmission}. Dashed gray lines denote the initial distribution while solid pink~(blue) lines the population of $\vert e_{+(-)} \rangle$ states after a time $t= 13 \gamma^{-1}$ has passed. The populations oscillate with a frequency given by the energy difference between upper and lower branches and signal the excitation of two modes. Backscattering is completely inhibited for the blue mode. As the disorder strength is increased from $\sqrt{W}/\Gamma_{0} =0, 0.625, 1.0$ between top and bottom pannels, the collective state of the regular chain losses coherence while that of the directional chain preserves it. The disorder strengths are to be compared with the transparency window of $2.5\Gamma_{0}$. } \label{Fig:localization} 
\end{center}
\end{figure}

Due to the asymetry in the dispersion relation of a non-reciprocal chain (see Fig.~\ref{Fig:transmission}) a state of well-defined wavevector finds less modes to backscatter to than one inside reciprocal chain, thus reducing the momentum spread. This is exemplified in Fig.~\ref{Fig:localization} where we plot the population of a spin-wave propagating inside an atomic chain for different disorder strengths and compare reciprocal and non-reciprocal responses. The spin-wave is again prepared in the state~(\ref{eq:spin_wave}). The plots show the distribution in position and reciprocal spaces after a time $t= 13 \Gamma_{0}^{-1}$ has passed with pink and blue lines used, respectively, for $\vert e_{+} \rangle$ and $\vert e_{-} \rangle$ polarizations. Notice first that when reciprocity is broken the superposition of the two excitation branches manifests as a beating in the population of $\vert e_{\pm} \rangle$ states, readily seen in reciprocal space. As the disorder strength is increased in panels~(b) and (c) the state begins to scatter into different quasimomentum components. In the reciprocal case the spread begins to occupy all the available states while in the non-reciprocal case there is only a small spread over the upper branch $\vert u\rangle $ where few modes are available. There is virtually no spread for the lower branch $\vert l \rangle $ as there are no modes available. 

The back and forward scattering eventually leads to localization of the excitation that prevents its transport~\cite{Anderson_1985}. For an atomic chain this localization describes a transient behavior: an excitation will eventually scatter out of the system through individual or collective channels. While the non-reciprocal chain has shown a reduced spread in momentum it arrives at the cost of a doubled radiation zone. It is found that, for the slow modes considered here, the loss is higher in the non-reciprocal case. This effect can be reduced for chains with a smaller lattice site.

\section{Conclusions}

In summary, we have presented a method to generate and probe the directional transport of excitations along an atomic chain. Directionality is achieved through an external control field that breaks the degeneracy between two excited states and induces a locally-varying dipole moment that follows a helical pattern, thus breaking time-reversal and parity symmetries. We find a simple formula where the probability for a free photon to enter the chain, propagate along collective decay channels, and then scatter out is readily calculated. This approach is based on detected events and has a direct connection to methods developed for electron transport in condensed matter physics~\cite{Landauer_1987}. We show that defect-induced backscattering is suppressed in directional chains, and the phase coherence between atoms of the chain survived for stronger disorder in comparison to regular chains. However, this comes at the cost of increased decay rate for strong disorder, due to the open nature of the system. 

\begin{acknowledgements}
We thank A.~Ortega and S.~Cardenas-Lopez for insightful discussions. R.G.-J. and A.~A.-G. acknowledge financial support by the National Science Foundation QII-TAQS (Award No. 1936359), and CAREER (Award No. 2047380). 
\end{acknowledgements}

%%%%%%%%%%%%%%%%%%%%%%%%%%%%%%%%%%%%

\end{document}